\documentclass[a4paper,12pt,twoside]{article}
\usepackage{myDiss}
\usepackage[bb=boondox]{mathalfa}
\usepackage{etoolbox}
\makeatletter
\def\cleardoublepage{\clearpage\if@twoside \ifodd\c@page\else
  \hbox{}
  \vspace*{\fill}
  \thispagestyle{empty}
  \newpage
  \if@twocolumn\hbox{}\newpage\fi\fi\fi}
\clearpage

\title{\Large{.}}
\author{\normalfont{S.~G.~Nagaraja, T.~Antretter and C.~Schuecker}}
\date{}
\begin{document}
\sloppy
\thispagestyle{empty}
\begin{center}
  {\Large\bfseries On the effect of isotropic and anisotropic
    dissipative response functions with associated and non-associated
    flow on the inelastic behaviour of polymeric composites.}
  \par\bigskip
  {S.~G.~Nagaraja$^{a,}$\footnote{corresponding author.~Tel.:+43 3842 402 4019
  \par\emph{E-mail address:}~swaroop.gaddikere-nagaraja@unileoben.ac.at
   \par\emph{URL:}~https://mechanik.unileoben.ac.at/},
  T.~Antretter$^a$ and C.~Schuecker$^b$}\par\medskip
  $^a$~Chair of Mechanics, Department of Physics, Mechanics and
  Electrical Engineering, \par
  Montanuniversitaet Leoben, Franz-Josef-Strasse 18/III, 8700 Leoben,
  Austria.\par\bigskip
  $^b$~Chair of Designing Plastics and Composite Materials, \par
  Department of Polymer Engineering and Science, Montanuniversitaet
  Leoben, \par
  Otto Gloeckel-Strasse 2, 8700 Loeben, Austria.
\end{center}
\pagenumbering{arabic}
\pagestyle{mainmatter} 
\renewcommand{\sectionmark}[1]{\markboth{#1}{}}
\begin{abstract}
  \noindent
  This article investigates the effect of using isotropic and anisotropic
plastic response functions in the analysis of the elastic-plastic
response of unidirectional fibre composites on the meso-scale. Three
model problems that use a Drucker-Prager-type pressure-dependent yield
function are considered to simulate the non-linearities exhibited by a
composite material. A further core ingredient is the analysis of a
canonical and non-conventional constitutive structure, with respect to
associated and non-associated flow response, where the use of latter is
motivated by the physical inconsistencies induced by the former under
shear dominated loads. These models are evaluated quantitatively by
comparison to experimental data. 
\par\noindent
\textbf{Keywords:} Anisotropy, plasticity, fibre-reinforced composites,
associative and non-associative flow rules.

\end{abstract}
\addtocontents{toc}{\protect\setcounter{tocdepth}{2}}
\msecta{Introduction.}
\label{intro}
Polymeric matrix composites have attracted wide attention in recent
years due to their superior mechanical properties such as strength,
stiffness and fatigue resistance, among others. They are increasingly
popular in the aerospace industry, automotive sector and civil
engineering applications. The growing demand for polymeric matrix
composites in these applications necessitates a comprehensive
understanding and accurate modelling of their complex behaviour under
various loading conditions. The predictive modelling of such composites
is thus essential for structural integrity, design and optimisation of
advanced engineering structures. Experimental investigations pertaining
to the non-linear behaviour of polymeric composites are documented in
\cite{weeks+sun1995,vogler+kyriakides1998,vogler+kyriakides1999}, among
others. It has been observed that the material response in the fibre
direction remains essentially elastic up to failure, whereas the
response to shear and transverse directions is non-linear and inelastic.
Recent research \cite{gsell+jacques+favre1990,gilat+goldberg+roberts2005} additionally implies that under shear dominated loads, considerable
irreversible strains develop that can be attributed to plasticity in the
matrix \cite{schuecker+pettermann2008}. As industries look for more
lightweight, sustainable and efficient materials, the demand for
thermoplastic matrix composites such as polyether ether ketone (PEEK) is
growing rapidly. Therefore, precise constitutive relations that account
for the elastic-plastic behaviour of polymeric composites are essential
to accurately predict the damage onset and failure of composite
materials. Theoretical frameworks for the description of anisotropic
plasticity with an emphasis on fibre-reinforced composites are well
established in the literature, see \cite{pettermann+planskeinsteiner+bohm+rammerstorfer1993,hsu+vogler+kyriakides1999,doghri+ouaar2003,doghri+adam+bilger2010} for the micro-mechanics based approach, \cite{hill1950,barlat+lege+brem1991,voyiadjis+thiagarajan1995,barlat+aretz+yoon+karabin+brem+dick2005,smith+liu+cao2015,nagaraja+pletz+schuecker2019,nagaraja+schuecker2019} for the continuum based approach and \cite{car+oller+onate2000,car+oller+onate2001} for alternative formulations using the concept
of mapped tensors. Though the micro-mechanical approach gives a better
understanding of reasons behind the experimentally observed behaviour,
it comes at a cost where higher number of coefficients are required for
the description of material response, namely the constituent properties
that are unavailable. Hence, attention is focussed on the latter two
approaches in the present work.

The essential aspect of modelling the elastic-plastic behaviour of
fibre-reinforced composites is the choice of the plastic response
functions, i.e., \emph{isotropic} or \emph{anisotropic} dissipative
functions. Isotropic dissipative functions provide a simplified analysis
and are computationally efficient \cite{sun+chen1989,chen+sun1993,xie+adams1995}, however they often lead to inaccurate predictions under
realistic and complex biaxial loading conditions. On the other hand,
anisotropic dissipative functions \cite{rogers1987,spencer1992} offer a
more realistic representation of the inelastic response under realistic
loading conditions \cite{tsai+sun2002,kontou+spathis2006,vyas+pinho+robinson2011,vogler+rolfes+camanho2013}. Despite this, the increased
computational complexity (higher number of coefficients which must be
obtained experimentally) poses significant challenges for practical
implementation. A further essential aspect is the selection of
\emph{flow rules} which play a crucial role in modelling the physical
behaviour of polymeric composites. Flow rules determine the direction of
the plastic deformation within the material \cite{miehe1998} and are
categorised as \emph{associative} and \emph{non-associative flow rules}.
Associative flow rules, also known as normality rules, enforce the
direction of plastic flow to be perpendicular to the yield surface. With
respect to polymeric composites, it is seen from \cite{vogler+rolfes+camanho2013,nagaraja+pletz+schuecker2019} that the predictions of
associative flow rules are in excellent agreement with the experimental
results under complex loading scenarios. Nevertheless, full complexity 
of the mechanical behaviour of the considered polymeric composite is not
reproduced. An argument in \cite{laux+gan+barton+thomsen2019} suggests
that a non-associative flow rule must be considered in order to
eliminate the physical inconsistency caused by the associated flow rules
under shear dominated loads, which has been investigated and reported in
the present work. A Non-associative flow rule relaxes the constraint
that the plastic flow direction must be normal to the yield surface
\cite{mosler+bruhns2009}. This flexibility offers additional features,
such as a purely deviatoric flow rule in presence of a
pressure-dependent yield function, to be incorporated in the modelling
framework.

The goal of this paper is to briefly outline a thermodynamically
consistent formulation of anisotropic plasticity for fibre-reinforced
composites, and investigate in detail the effect of aforementioned
modelling choices on the non-linear inelastic behaviour of composites.
In a first step, elements of infinitesimal plasticity theory are
introduced that builds the necessary background for the subsequent
derivations. This is followed by a brief discussion on the general form
of the energetic and dissipative response functions based on
representation theorems. Next, three continuum based models are
presented to simulate the non-linearities exhibited by the composite.
The first model is a modified Drucker-Prager-type model, formulated
following \cite{xie+adams1995}, in which the classical isotropic
Drucker-Prager-type pressure-dependent isotropic yield criterion is
modified for use with unidirectional composites. This is followed by a
second model, which is a modified version of the model proposed by Car,
Oller and O\~nate \cite{car+oller+onate2000,car+oller+onate2001}. It
assumes the existence of a fictitious isotropic space where a mapped
problem is solved. The third model is an extension of Model-I into an
\emph{anisotropic} form using representation theorems, discussed in
\cite{nagaraja+pletz+schuecker2019,nagaraja+schuecker2019}. A further
key aspect is the qualitative and quantitative evaluation of the
aforementioned models by comparison to experimental data. Concluding
remarks appear in the end.

\msecta{Fundamentals of infinitesimal strain plasticity.}
\label{fundamentals}
The infinitesimal plasticity theory \cite{papadopoulos+taylor1994,simo+hughes2000}, based on the additive decomposition of the total strain into
elastic and plastic parts, allows for the existence of a symmetric
second-order plastic strain tensor $\STRAIN^p$ as an internal variable.
In addition, it also allows for the existence of hardening variables,
which in the present work are a symmetric second-order tensor $\Balpha$
and a scalar $\alpha$, characterising kinematic and isotropic hardening
respectively. These variables allow for the definition of a
scalar-valued energetic potential as
\begin{equation}
  \psi = \psi(\STRAIN-\STRAIN^p,\Balpha,\alpha)
  = \frac{1}{2}{\Norm{\STRAIN-\STRAIN^p}}^2_{\STIFFNESSMATRIX}
  + \frac{1}{2}{\Norm{\Balpha}}^2_{\nIH}
  + \frac{h}{n+1}(\bar\alpha+\alpha)^{n+1}
  \ ,
  \label{general_energy}
\end{equation}
where $\Norm{\cdot}^2_{(\STIFFNESSMATRIX,\nIH)}= \left<\cdot,\cdot\right>_{(\STIFFNESSMATRIX,\nIH)}$, with $\left<\cdot,\cdot\right>$ denoting the
inner product, $\STRAIN=\nabla_{\sym}\Bu$ is the total strain tensor
defined by the symmetric part of the displacement gradient,
$\STIFFNESSMATRIX$ and $\nIH$ are fourth-order symmetric anisotropic
elastic modulus and kinematic hardening modulus tensors, respectively.
The parameters $h$ and $n$ denote the isotropic hardening modulus and
exponent, respectively, while $\bar\alpha$ denotes prestrain which is
necessary for numerical reasons and is set to a very low value such that
it has negligible effect on the results \cite{nagaraja+pletz+schuecker2019}.

The closed form expressions for the stress tensor and driving forces
associated with the potential $\psi$ are obtained by the Coleman-Noll
argument \cite{coleman+noll1963,coleman+gurtin1967} as
\begin{equation}
  \begin{array}{l@{\ }c@{\ }l}
    \disp\STRESS &=& \disp +\partial_{(\STRAIN-\STRAIN^p)}\psi
    = \disp\STIFFNESSMATRIX:(\STRAIN-\STRAIN^p) \ , \\[2mm]
    \disp\Bbeta &=& \disp -\partial_{\Balpha}\psi = \disp-\nIH:\Balpha
    \ , \\[2mm]
    \disp \beta &=& \disp -\partial_{\alpha}\psi =-h(\bar\alpha+\alpha)^n
    \ .
  \end{array}
  \label{general_stress}
\end{equation}
Next, as a main characteristic of the elastic-plastic material response,
an elastic domain $\elasticdomain$ is assumed, defined by
\begin{equation}
  \elasticdomain =
  \{(\STRESS,\Bbeta,\beta)\in\elR^{6+6+1}|\chi(\STRESS,\Bbeta,\beta)
  \leq 0\}\ ,
  \label{elastic-domain}
\end{equation}
where $\chi = \chi(\STRESS,\Bbeta,\beta)$ is the yield function in the 
space of admissible driving forces. The yield function is of the
generalised Drucker-Prager-type, and takes the form
\begin{equation}
  \chi = \kappa{p}
       + \Norm{\BSigma}_{\nIP}
       - \sqrt{\frac{2}{3}}[y_0 - \beta]
       + \frac{b}{2}{\Norm{\Bbeta}}^2_\nIQ
       \WITH \BSigma=\STRESS+\Bbeta \ ,
  \label{general_yield}
\end{equation}
where $\Norm{\cdot}_{\nIP}= \sqrt{\left<\cdot,\cdot\right>_{\nIP}}$,
$\kappa$ is the coefficient of the hydrostatic pressure,
$p = \frac{1}{3}\tr[\BSigma]$ is the hydrostatic pressure in terms of
the effective stress tensor $\BSigma$, $y_0$ characterises the initial
threshold yield stress, $b$ governs the non-linearity of kinematic
hardening, and $\{\nIP,\nIQ\}$ are symmetric fourth-order deviatoric
Hill-type tensors.

Taking into account \reqs{general_energy} and \eqref{general_yield}, the
generalised normality condition \cite{lubliner1997,khan+huang1995,aldakheel+miehe2017} yields the flow rule and rate equations for the hardening
variables as
\begin{equation}
  \begin{array}{l@{\ }c@{\ }l}
    \disp\dot{\STRAIN}^p &=& \disp\lambda\partial_{\STRESS}\chi
      = \disp\lambda\left\{\frac{\kappa}{3}\Bone
        + \frac{\nIP:\BSigma}{\Norm{\BSigma}_{\nIP}}\right\} \\[5mm]
    \disp\dot{\Balpha} &=& \lambda\partial_\Bbeta\chi
      = \disp\lambda\left\{\frac{\kappa}{3}\Bone
       +\frac{\nIP:\BSigma}{\Norm{\BSigma}_{\nIP}} + b\nIQ:\Bbeta\right\}
\\[5mm]
   \disp\dot{\alpha} &=& \disp\lambda\partial_{\beta}\chi
   = \disp\lambda\sqrt{\frac{2}{3}}
   = \disp\sqrt{\frac{2}{3}}\Norm{\dot{\STRAIN}^p}
  \end{array}\ ,
  \label{general_flowrule}
\end{equation}
where $\lambda$ denotes the amount of the plastic flow and $\Bone$ is
the second-order identity tensor. Additionally, the rate equations
\eqref{general_flowrule} are supplemented by Karush-Kuhn-Tucker-type
loading-unloading conditions such that
\begin{equation}
  \lambda \geq 0,\;\;\chi \leq 0 \AND \lambda\chi=0 \ ,
  \label{general_kkt}
\end{equation}
from which the amount of the plastic flow $\lambda$ can be computed. In
view of the fact that the rates of the internal variables are normal to
the yield surface ($\chi=0$), evolution laws in \req{general_flowrule}
are referred to as associated flow rules.

For a non-associated flow response, where the canonical normal
directions of the evolution equations \eqref{general_flowrule} do not
characterise the real material response, the constitutive response is
modified by introducing an additional function $\phi$, henceforth
referred to as the plastic flow potential. It is assumed to be of the
same form as the yield function but with a different set of governing
material coefficients \cite{mosler+bruhns2009} such that
\begin{equation}
  \phi = \tilde\kappa{p}
       + \Norm{\BSigma}_{\nIP}
       - \sqrt{\frac{2}{3}}[\tilde{y}_0 - \beta]
       + \frac{\tilde b}{2}{\Norm{\Bbeta}}^2_\nIQ \ ,
  \label{general_flow}
\end{equation}
where $\tilde\kappa,\tilde{y}_0,\tilde{b}\neq\kappa,y_0,b$.
Consequently, the flow rule and the rate equations for the hardening
variables take the form
\begin{equation}
  \begin{array}{l@{\ }c@{\ }l}
    \disp\dot{\STRAIN}^p &=& \disp\lambda\partial_{\STRESS}\phi
      = \disp\lambda\left\{\frac{\tilde\kappa}{3}\Bone
        + \frac{\nIP:\BSigma}{\Norm{\BSigma}_{\nIP}}\right\}\\[5mm]
    \disp\dot{\Balpha} &=& \disp\lambda\partial_\Bbeta\phi
      = \disp\lambda\left\{\frac{\tilde\kappa}{3}\Bone
      + \frac{\nIP:\BSigma}{\Norm{\BSigma}_{\nIP}}
      + \tilde{b}\nIQ:\Bbeta\right\}
\\[5mm]
    \disp\dot{\alpha} &=& \disp\lambda\partial_{\beta}\phi
    = \disp\lambda\sqrt{\frac{2}{3}}
  \end{array} \ .
  \label{general_flowrule-NA}
\end{equation}
Note that \req{general_flowrule-NA} replaces the normality rules in
\req{general_flowrule} though the loading-unloading conditions remain
unchanged.
\begin{remark}
The framework of infinitesimal plasticity discussed so far, assumes a
\emph{rate-independent} setting such that the material behaviour will
not qualitatively change for varying load rates $\dot\STRAIN$. To
account for rate-dependency, one can use the classical
\emph{rate-dependent} formulation of Perzyna-type \cite{perzyna1971,
lubliner1972}, which yields the plastic multiplier as
\begin{equation}
  \lambda = \frac{1}{\eta}(\chi^+) \WITH
  (\chi^+) := \frac{1}{2}\big(\chi + \Abs{\chi}\big) \ ,
  \label{amount-perzyna}
\end{equation}
where $\eta\in(0,\infty)$ is the penalty parameter that characterises
time-dependent viscous plastic flow and $(\chi^+)$ denotes a
monotonically increasing ramp function \cite{miehe+apel+lambrecht2002,papadopoulos+taylor1994}. Equation \eqref{amount-perzyna} is known as the
\emph{pseudo-consistency condition} which yields the rate-dependent
$\lambda$ in terms of viscosity $\eta$ and the ramp function $(\chi^+)$.
The strain-rate sensitivity of polymeric composites, particularly the
shear and compressive response of AS4/PEEK is documented in \cite{vogler+kyriakides1999}, where it is seen that the change in elastic modulus is
relatively small, whereas, for the inelastic regime, both the shear and
compressive stress increases significantly with the strain rate. This
phenomenon can be captured with the aid of \req{amount-perzyna} for the
present case. A rate-dependent viscoplastic formulation, cf.~\cite{nagaraja+schuecker2019}, offers additional flexibility for the choice of the
slope ($\kappa/\tilde\kappa$) of pressure dependency, which should be
rather low as pointed out in \cite{nagaraja+pletz+schuecker2019}.
\end{remark}

\msecta{Transversely isotropic system $\symG$ generated by
  ($\BQ_{\parallel\Ba},\BQ^{\pi}_{\perp\Ba}$).}
\label{invariant}
In this section, an explicit form of the scalar-valued energetic
potential and the yield function is derived for the transversely
isotropic symmetric group with the aid of representation theorems
\cite{boehler1979}. From a continuum viewpoint, a composite is usually
characterised by two different symmetry groups based on their inherent
micro-structure. If the material is reinforced by fibres in one
direction, then the composite has only a single preferred direction and
is characterised by the \emph{transversely isotropic} symmetry group.
Typical example is a unidirectional fibre-reinforced composite. It is
also conceivable for a composite material to be reinforced by fibres in
more than one direction, such as a woven fabric that has fibres aligned
in two perpendicular directions. Such materials belong to the
\emph{orthorhombic} symmetry group and are characterised by the
existence of two preferred directions. Focus is purely restricted to the
former symmetry group in the present work. In this regard, let $\Ba$ be
a positively oriented vector denoting the preferred direction with
$\Norm{\Ba} = 1$. The considered symmetry group $\symG$ is generated by
the orthogonal tensors $\BQ_{\parallel\Ba}$ and $\BQ^{\pi}_{\perp\Ba}$ which
correspond to arbitrary rotations relative to the vector $\Ba$ and about
a vector perpendicular to $\Ba$ by the angle $\pi$ respectively, see
\cite{lu+zhang2005}. A key approach to the formulation of constitutive
response functions with the aid of representation theorems is the
construction of isotropic tensor functions with an extended set of
arguments, known as structural tensors \cite{liu1982}. Recall here, that
the transversely isotropic symmetry group is fully characterised by a
single symmetric second-order structural tensor $\Bm$, defined by
\begin{equation}
  \Bm = \Ba\otimes\Ba\ ,
  \label{m-tensor}
\end{equation}
where $\Bm$ is invariant to rotations $\BQ$ of the symmetry group
$\symG$, i.e.~$\BQ\Bm\BQ^T = \Bm\;\;\forall\;\;\BQ\in\symG$, see
\cite{zheng+spencer1993,zheng1994} for details. Appealing to the
representation theorems for isotropic scalar and tensor functions of
two symmetric second-order tensors $\BA$ and $\Bm$, an irreducible
integrity basis for the transversely isotropic symmetry group is given
by
\begin{equation}
  \calI = \left\{\tr[\BA],\tr\left[\BA^2\right],\tr\left[\BA^3\right],
  \tr[\Bm\BA],\tr\left[\Bm\BA^2\right]\right\} \ ,
  \label{tsiso_intbasis-1}
\end{equation}
see \cite{smith1965,spencer1971,spencer1987}. Following \cite{nagaraja+pletz+schuecker2019,nagaraja+schuecker2019}, the set of invariants in
\req{tsiso_intbasis-1} can be reformulated to
\begin{equation}
  \tilde{\calI} = \left\{\tilde{I}_1,\cdots,\tilde{I}_5\right\}
                = \left\{\tr[\Bm\BA],\tr[(\Bone-\Bm)\BA],
                         \tr\left[\Bm\BA^2\right],
  \tr\left[\left(\frac{1}{2}\Bone-\Bm\right)\BA^2\right],\det[\BA]
  \right\}\ ,
  \label{tsiso_intbasis-2}
\end{equation}
where the first two invariants in \req{tsiso_intbasis-2} are normal
modes and the next two modes are shear. Scalar-valued constitutive
functions can now be constructed by taking combinations of the
invariants defined above. In particular, a \textit{quadratic} potential
$\Pi$ can be written as
\begin{equation}
  \Pi = \frac{\mu_1}{2}{\tilde{I}_1}^2 + \frac{\mu_2}{2}{\tilde{I}_2}^2
  + \mu_3 \tilde{I}_1\tilde{I}_2 + 2\mu_4\tilde{I}_3 + 2\mu_5\tilde{I}_4
  \ ,
  \label{INV_energy}
\end{equation}
where $\mu_{1-5}$ are five independent Lam\'e parameters required to
describe the transversely isotropic response. Note that the cubic
invariant $\tilde{I}_5$ is neglected as it is most suitable for
modelling metal plasticity. The closed form expression of the
fourth-order Hessian associated with the potential reads
\begin{equation}
  \begin{array}{r@{\ }r@{\ }l}
    \disp \nIA = \disp\Pi,_{\BA\BA}
    &=& \mu_1\Bm\otimes\Bm
    + \mu_2\big\{(\Bone-\Bm)\otimes(\Bone-\Bm)\big\}\\[1mm]
    &+& \disp\mu_3\big\{
    \Bm\otimes(\Bone-\Bm) + (\Bone-\Bm)\otimes\Bm\big\}\\[1mm]
    &+& \disp\mu_4\big\{
    (\Bone\oplus\Bm) + (\Bone\ominus\Bm) +
    (\Bm\oplus\Bone) + (\Bm\ominus\Bone)\big\}\\[1mm]
    &+& \disp\mu_5\big\{(\Bone\oplus\Bone) + (\Bone\ominus\Bone)
    -(\Bone\oplus\Bm) - (\Bone\ominus\Bm) -
    (\Bm\oplus\Bone) - (\Bm\ominus\Bone)\big\}
    \label{INV_emoduli}
  \end{array}\ ,
\end{equation}
where the identities
\begin{equation}
  \begin{array}{r@{\ }l@{\ }l}
   \disp\big\{(\bullet)\otimes(\bullet)\big\}_{ijkl} &=&
   \disp(\bullet)_{ij}(\bullet)_{kl} \\[1mm]
   \disp\big\{(\bullet)\oplus(\bullet)\big\}_{ijkl} &=&
   \disp(\bullet)_{ik}(\bullet)_{jl} \\[1mm]
   \disp\big\{(\bullet)\ominus(\bullet)\big\}_{ijkl} &=&
   \disp(\bullet)_{il}(\bullet)_{jk}
  \end{array} \ ,
\label{INV_emoduli-1}
\end{equation}
have been introduced. For the choice of $\Ba=[1,0,0]^T$, the Hessian
$\nIA$ appears in the coordinate form
\begin{equation}
  \begin{array}{c@{\ }c@{\ }c}
    \disp [\nIA] =
    \begin{bmatrix}
      \disp \mu_1+4\mu_4-2\mu_5 & \mu_3 & \mu_3 & 0 & 0 & 0 \\[1mm]
      \disp            & \mu_2+2\mu_5 & \mu_2 & 0 & 0 & 0 \\[1mm]
      \disp            &            & \mu_2+2\mu_5 & 0 & 0 & 0 \\[1mm]
      \disp            &            &            & \mu_4 & 0 & 0 \\[1mm]
      \disp            &            &            &  & \mu_4 & 0 \\[1mm]
      \disp \sym       &            &            &  &  & \mu_5
    \end{bmatrix}
  \end{array} \ ,
  \label{Tsiso-Hessian}
\end{equation}
where it is seen that the fourth-order tensor $\nIA$ satisfies major and
minor symmetries, i.e.,
\begin{equation}
  \nIA_{ijkl} = \nIA_{jikl} = \nIA_{ijlk}
  = \nIA_{klij} \ .
  \label{A_symm}
\end{equation}

\msecta{Continuum formulation.}
\label{models}
In this section, simple models of anisotropic plasticity are discussed
which can be used for the analysis of infinitesimal elastic-plastic
deformation of fibre-reinforced composites. Starting with the elastic
response functions, use is made of \req{Tsiso-Hessian} to first define
the scalar-valued energetic function. Based on the elements of
infinitesimal plasticity theory introduced in \rSec{fundamentals}, three
models chosen from the literature are presented to simulate the
non-linearities exhibited by the composite. The first model is a
modified Drucker-Prager model (Model-I), formulated following \cite{xie+adams1995}, in which the classical Drucker-Prager-type
pressure-dependent isotropic yield criterion is modified for use with
fibre-reinforced composites. Here, only the volumetric-isochoric
decomposition of the stress tensor is considered. This is followed by a
second model, which is a modified version of the model proposed by Car,
Oller and O\~nate \cite{car+oller+onate2000,car+oller+onate2001}
(Model-II). It assumes the existence of a fictitious isotropic space
where a mapped problem is solved. The third model (Model-III) is an
extension of Model-I into an \emph{anisotropic} form using
representation theorems, discussed in \cite{nagaraja+pletz+schuecker2019,nagaraja+schuecker2019}. All the models use the classical
Drucker-Prager-type yield function, mainly for the computational aspects
owing to its smooth surface and numerical stability.
\msectb{Elastic response functions.}
\label{ERF}
Setting $\nIA=\nIE$ in \req{general_energy}, the stress tensor defined
in \reqi{general_stress}{1}, is obtained in terms of the respective
governing coefficients, see also \cite{miehe+apel+lambrecht2002}. Note
that the Lam\'e parameters are identified in terms of the corresponding
engineering constants using the prescription suggested in \cite{schroder+gruttmann+loblein2002}. For the sake of simplicity and absence of
relevant experimental data, $\nIH=\mathbb{0}$ is reasonably assumed.
Next, the plastic response functions are formulated with the preceding
definitions at hand.
\msectb{Plastic response functions.}
\label{PRF}
In the context of formulating the plastic response functions for
unidirectional fibre-reinforced composites, Spencer \cite{spencer1992}
introduced a physically motivated plasticity inducing stress tensor
which is obtained in an additive format from the overall stress, the
hydrostatic pressure and the deviatoric fibre stress for a given fibre
direction $\Bm$ \cite{spencer1992,rogers1987,lu+zhang2005}. This is
specified to the present model problems by defining a fourth-order
projection tensor
\begin{equation}
  \nIP = (\Bone\oplus\Bone) + (\Bone\ominus\Bone)
  - \frac{1}{3}(\Bone\otimes\Bone)
  - \frac{3}{2}(\Bm^{\prime}\otimes\Bm^{\prime})
  \WITH \Bm^{\prime} = \Bm - \frac{1}{3}\Bone \ ,
  \label{stress_projection}
\end{equation}
with the following characteristics
\begin{equation}
  \tr[\nIP:\Bone] = 0 \AND \tr[\nIP:\Bm] = 0 \ .
  \label{stress_pprojection-2}
\end{equation}
Additionally, let $p$ denote the hydrostatic pressure such that
\begin{equation}
  p = \frac{1}{3}\tr[(\Bone-\Bm)\BSigma]\ .
  \label{stress_hydro}
\end{equation}
Note, $\nIP$ and $p$ in \reqs{stress_projection} and 
\eqref{stress_hydro}, respectively, ensure a linear elastic fibre
response.
\msectc{Model-I.}
\label{M1}
Appealing to \reqs{stress_projection} and \eqref{stress_hydro}, the
yield function for Model-I now reads
\begin{equation}
  \chi = \kappa{p}
       + \Norm{\BSigma}_{\nIP}
       - \sqrt{\frac{2}{3}}[y_0 - \beta]
       + \frac{b}{2}{\Norm{\Bbeta}}^2_\nIQ \ .
  \label{yield-1}
\end{equation}
The normality rules in \req{general_flowrule} reformulate with \req{yield-1} to
\begin{equation}
  \begin{array}{l@{\ }c@{\ }l}
    \disp\dot{\STRAIN}^p &=&
        \disp\lambda\left\{\frac{\kappa}{3}(\Bone-\Bm)
      + \frac{\nIP:\BSigma}{\Norm{\BSigma}_{\nIP}}\right\} \\[5mm]
    \disp\dot{\Balpha} &=&
    \disp\lambda\left\{\frac{\kappa}{3}(\Bone-\Bm)
  + \frac{\nIP:\BSigma}{\Norm{\BSigma}_{\nIP}} + b\nIQ:\Bbeta\right\}
\\[5mm]
    \disp\dot{\alpha} &=& \disp\lambda\sqrt{\frac{2}{3}}
  \end{array} \ .
  \label{Y1_flowrule}
\end{equation}
For the case of non-associative plasticity, the corresponding flow rule
and rate equations for the hardening variables can be obtained in a
similar manner. In particular, a deviatoric flow potential is chosen
following \cite{mosler+bruhns2009} as
\begin{equation}
  \phi := \chi\big\vert_{\kappa=0}
        = \Norm{\BSigma}_{\nIP}
        - \sqrt{\frac{2}{3}}[y_0 - \beta]
        + \frac{b}{2}{\Norm{\Bbeta}}^2_\nIQ \ ,
  \label{flowpotential-1}
\end{equation}
based on which the evolution equations in \req{general_flowrule-NA} take
the form
\begin{equation}
  \begin{array}{l@{\ }c@{\ }l}
    \disp\dot{\STRAIN}^p &=&
    \disp\lambda\left\{\frac{\nIP:\BSigma}{\Norm{\BSigma}_{\nIP}}\right\}
\\[5mm]
  \disp\dot{\Balpha} &=&
  \disp\lambda\left\{\frac{\nIP:\BSigma}{\Norm{\BSigma}_{\nIP}}
                    + b\nIQ:\Bbeta\right\}
\\[5mm]  
  \disp\dot{\alpha} &=& \disp\lambda\sqrt{\frac{2}{3}}
  \end{array} \ .
  \label{Y1_flowrule-NA}
\end{equation}
\msectc{Model-II.}
\label{M2}
This is a material model for fibre-reinforced composites based on the
work of Car, Oller and O\~nate \cite{car+oller+onate2000,
car+oller+onate2001}. It assumes the existence of a fictitious isotropic
space where a mapped problem is solved. The real and fictitious spaces
are related by means of fourth-order transformation tensors which are
formulated based on the available information of strengths in the
respective spaces. The real anisotropic space is regarded as a
homogenised composite material, while the fictitious isotropic space
characterises the matrix material to which plasticity is usually
restricted.

Let $\BnY$ and $\overline\BnY$ each represent a second-order yield
strength tensor for the real anisotropic space and the fictitious
isotropic space, respectively. Based on the yield strength tensors, a
fourth-order space transformation tensor for the stress is proposed as
\begin{equation}
  \nIM = \frac{1}{2}\big\{(\overline\BnY\oplus\BnY^{-1})
  + (\BnY^{-1}\oplus\overline\BnY)\big\} \ ,
  \label{plasticity_iso-MP2-_trans_tensor}
\end{equation}
which satisfies the major and minor symmetries
\begin{equation}
  \nIM_{ijkl} = \nIM_{jikl} = \nIM_{ijlk} = \nIM_{klij} \ .
  \label{plasticity_iso-MP2-_symm}
\end{equation}
The fourth-order tensor $\nIM$ relates the stress tensor and the
back-stress tensor in the real and fictitious spaces as
\begin{equation}
  \overline\STRESS = \nIM:\STRESS \AND
  \overline\Bbeta = \nIM:\Bbeta \ ,
  \label{plasticity_iso-MP2-_stress_trans}
\end{equation}
with $\STRESS$ denoting the stress tensor in the real anisotropic space,
defined in \reqi{general_stress}{1}. It should be noted here that Car,
Oller and O\~nate \cite{car+oller+onate2000,car+oller+onate2001} defined
the transformation tensor as $\nIM=\overline\BnY\otimes\BnY^{-1}$, but it
is slightly modified in the present work to get a compact representation
of the transformation tensor. In what follows, the quantities $(\cdot)$
and $\overline{(\cdot)}$ relate to the real anisotropic and the
fictitious isotropic space, respectively. Analogous to \req{plasticity_iso-MP2-_stress_trans}, the relationship between the elastic strain in
both spaces is defined by
\begin{equation}
  (\overline\STRAIN - \overline\STRAIN^p) = 
  \nIN:(\STRAIN - \STRAIN^p) \ ,
  \label{plasticity_iso-MP2-_strain_trans}
\end{equation}
which implies the non-uniqueness of elastic strain during space
transformation. The fourth-order strain transformation tensor
$\nIN$ is obtained with the aid of \req{plasticity_iso-MP2-_stress_trans} as
\begin{equation}
  \nIN = \overline\STIFFNESSMATRIX^{-1}:\nIM:\STIFFNESSMATRIX \ ,
  \label{plasticity_iso-MP2-_trans_tensor-1}
\end{equation}
where $\STIFFNESSMATRIX$ and $\overline\STIFFNESSMATRIX$ are the elastic
modulus tensors in the real anisotropic and fictitious isotropic spaces,
respectively. The fourth-order tensor $\STIFFNESSMATRIX$ includes the
actual properties of the material, i.e., \req{Tsiso-Hessian}, whereas
the choice of $\overline\STIFFNESSMATRIX$ can be mathematically
arbitrary \cite{car+oller+onate2000,car+oller+onate2001} but should
physically represent the matrix constituent. In what follows, the
governing constitutive equations of the plastic deformation process are
specified in the fictitious isotropic space. Note that it is equivalent
to formulate the model in either of the two spaces because of the
invariance of the dissipation postulate \cite{car+oller+onate2000}. Due
to the advantages of the existing algorithms for isotropy, modelling in
the fictitious isotropic space is considered here.

Starting from \req{general_yield}, the yield function in the fictitious
isotropic space is given by
\begin{equation}
  \chi = \kappa\overline{p} +  \Norm{\overline\BSigma}_{\overline\nIP}
  - \sqrt{\frac{2}{3}}[y_0 - \beta]
  + \frac{b}{2}\Norm{\overline\Bbeta}^2_{\overline{\nIQ}} \WITH
  \overline\BSigma=\overline\STRESS+\overline\Bbeta\ ,
  \label{plasticity_iso-MP2-_yield}
\end{equation}
where $\overline{p}=\frac{1}{3}\tr[\overline\BSigma]$ is the hydrostatic
pressure, and $\overline\nIP$ is the symmetric isotropic fourth-order
deviatoric projection tensor given by
\begin{equation}
  \overline\nIP = (\Bone\oplus\Bone) + (\Bone\ominus\Bone)
  - \frac{1}{3}(\Bone\otimes\Bone) \ .
  \label{plasticity_iso-MP2-_stress_pind}
\end{equation}
Recall here that the deviatoric fourth-order Hill-type tensor
$\overline\nIQ$ is similar to $\overline\nIP$, and governs the
non-linearity of kinematic hardening. The flow rule and rate equations
of the hardening variables within the isotropic space are specified
analogous to \req{Y1_flowrule} as
\begin{equation}
  \begin{array}{l@{\ }c@{\ }l}
   \disp\dot{\overline\STRAIN}^p &=&
   \disp\lambda\left\{\frac{\kappa}{3}\Bone
  + \frac{\overline\nIP:\overline\BSigma}
  {\Norm{\overline\BSigma}_{\overline\nIP}}\right\} \\[5mm]
  \disp\dot{\overline\Balpha} &=&
  \disp\lambda\left\{\frac{\kappa}{3}\Bone
  + \frac{\overline\nIP:\overline\BSigma}
         {\Norm{\overline\BSigma}_{\overline\nIP}}
  + b\overline{\nIQ}:\overline\Bbeta\right\} \\[5mm]
  \disp\dot\alpha &=& \disp\lambda\sqrt{\frac{2}{3}}
  \label{plasticity_iso-MP2-_flowrule}
  \end{array} \ .
\end{equation}
Furthermore, a separate deviatoric flow potential that governs
the evolution of plastic variables within the framework of
non-associative plasticity is defined as
\begin{equation}
  \phi := \chi\big\vert_{\kappa=0}
        = \Norm{\overline\BSigma}_{\overline\nIP}
        - \sqrt{\frac{2}{3}}[y_0 - \beta]
        + \frac{b}{2}\Norm{\overline\Bbeta}^2_{\overline\nIQ} \ ,
\label{plasticity_iso-MP2-_flow}
\end{equation}
based on which the evolution equations
\eqref{plasticity_iso-MP2-_flowrule} reformulate respectively to
\begin{equation}
  \begin{array}{l@{\ }c@{\ }l}
   \disp\dot{\overline\STRAIN}^p &=&
   \disp\lambda\left\{\frac{\overline\nIP:\overline\BSigma}
  {\Norm{\overline\BSigma}_{\overline\nIP}}\right\} \\[5mm]
  \disp\dot{\overline\Balpha} &=&
  \disp\lambda\left\{\frac{\overline\nIP:\overline\BSigma}
         {\Norm{\overline\BSigma}_{\overline\nIP}}
  + b\overline{\nIQ}:\overline\Bbeta\right\} \\[5mm]
  \disp\dot\alpha &=& \disp\lambda\sqrt{\frac{2}{3}}
  \label{plasticity_iso-MP2-_flowrule-1}
  \end{array} \ .
\end{equation}
for the non-associative flow response. Equations \eqref{plasticity_iso-MP2-_yield}, \eqref{plasticity_iso-MP2-_flowrule} and \eqref{plasticity_iso-MP2-_flowrule-1} are solved by a general elastic predictor-plastic
corrector algorithm described in \cite{simo+hughes2000,schroder+gruttmann+loblein2002}, which gives the consistent update of the stress tensor,
plastic strain tensor, hardening variables and the algorithmically
consistent elastic-plastic tangent modulus. With these tensorial
quantities in the fictitious isotropic space at hand, the corresponding
real anisotropic counterparts are obtained by a straightforward
transformation as follows
\begin{equation}
  \STRESS^{\ep} = \nIM^{-1}:\overline\STRESS^{\ep} \AND
  \STIFFNESSMATRIX^{\ep} = \nIM^{-1}:\overline\STIFFNESSMATRIX^{\ep}:\nIN
  \ .
  \label{plasticity_iso-MP2-_back_trans}
\end{equation}
In summary, Model-II requires the following material properties to
describe the elastic-plastic response of unidirectional fibre-reinforced
composite materials:
\begin{itemize}
\item Real anisotropic space:
  \begin{itemize}
  \item elastic parameters $\mu_{1-5}$,
  \item yield strength tensor $\BnY$.
  \end{itemize}
\item Fictitious isotropic space:
  \begin{itemize}
  \item plastic parameters $\kappa$, $y_0$, $h$, $\overline\alpha,n$ and
  $b$,
  \item yield strength tensor $\overline\BnY$.
  \end{itemize}
\end{itemize}
\msectc{Model-III.}
\label{M3}
The third model (Model-III) considers an extension of the isotropic
plastic response functions of Model-I into \emph{anisotropic} forms
using the representation theorems discussed in \cite{nagaraja+pletz+schuecker2019,nagaraja+schuecker2019}. Precisely, a further decomposition
of $\nIP$ in \req{stress_projection} into the two shear modes associated
with the symmetry group is considered. To this end, we define
\begin{equation}
\nIP = \nIP_1 + \nIP_2 \ ,
\label{M2-projection-1}
\end{equation}
where, following \cite{lu+zhang2005}, it can be verified that
\begin{equation}
\nIP_1 = \frac{1}{2}[(\Bone\oplus\Bm) + (\Bm\oplus\Bone)
                    +(\Bone\ominus\Bm) + (\Bm\ominus\Bone)
                    - 2(\Bm\otimes\Bm)] \AND
\nIP_2 = \nIP - \nIP_1 \ .
\label{M2-projection-2}
\end{equation}
In the equation above $\nIP_1$ and $\nIP_2$ are the in-plane and
transverse shear modes. The fourth-order tensors $\nIQ_1$ and $\nIQ_2$
are obtained similar to $\nIP_1$ and $\nIP_2$, respectively. Equation
\eqref{M2-projection-2} allows for the definition of the yield function
for Model-III, entirely analogous to \cite{nagaraja+pletz+schuecker2019,nagaraja+schuecker2019} as
\begin{equation}
  \chi = \kappa{p}
       + \Norm{\BSigma}_{(a_1\nIP_1+a_2\nIP_2)}
       - \left[1 - \frac{\beta}{y_{12}}\right]
       + \frac{1}{2}{\Norm{\Bbeta}}^2_{(b_1\nIQ_1+b_2\nIQ_2)} \ ,
  \label{yield-2}
\end{equation}
see also \cite{papadopoulos+lu2001}, where $p$ is the hydrostatic
pressure defined in \req{stress_hydro}. The four parameters
$\kappa,a_1,a_2$ and $y_{12}$ in the equation above govern the
transversely isotropic plastic yielding. They are determined by the
evaluation of the yield function \eqref{yield-2} for two simple shear
tests and one normal (compression) test, with $\beta=0$ and
$\Bbeta=\Bzero$ \cite{nagaraja+pletz+schuecker2019}, as
\begin{equation}
  \kappa = \frac{1}{\sqrt{2}y_{23}} - \frac{1}{y_{22c}} \ , \quad
  a_1 = \frac{1}{y_{12}^2} \AND
  a_2 = \frac{1}{y_{23}^2} \ .
  \label{yield-2-par}
\end{equation}
where $y_{12},y_{23}$ and $y_{22}$ denote the in-plane, transverse shear
and transverse compressive yield stress respectively. The remaining two
parameters $b_1$ and $b_2$ in \req{yield-2} govern the non-linearity of
kinematic hardening. Considering the fact that the Hessian of the
squared Euclidean norm is the identity matrix which is axiomatically
positive definite, the requirement for convexity of the yield surface
\eqref{yield-2} is given by $a_{1-2}\geq 0$, see also Naghdi-Trapp
inequality \cite{naghdi+trapp1975-1,naghdi+trapp1975-2,casey1984}, which
is generally fulfilled by \req{yield-2-par}.

The normality rules for Model-III follow with the aid of \req{yield-2}
as
\begin{equation}
  \begin{array}{l@{\ }l@{\ }l}
    & \disp\dot{\STRAIN}^p &
    = \disp\lambda\left\{\frac{\kappa}{3}(\Bone-\Bm)
    + \frac{(a_1\nIP_1+ a_2\nIP_2):\BSigma}
           {\Norm{\BSigma}_{(a_1\nIP_1+a_2\nIP_2)}}\right\}\\[6.5mm]
    & \disp\dot{\Balpha} &
    = \disp\lambda\left\{\frac{\kappa}{3}(\Bone-\Bm)
    + \frac{(a_1\nIP_1+ a_2\nIP_2):\BSigma}
           {\Norm{\BSigma}_{(a_1\nIP_1+a_2\nIP_2)}}
    + (b_1\nIQ_1+ b_2\nIQ_2):\Bbeta\right\} \\[6.5mm]
    & \disp \dot{\alpha} &= \disp\lambda\frac{1}{y_{12}}
\end{array} \ .
  \label{evolution-M2}
\end{equation}
Analogous to the previous two models, a plastic flow potential that
governs the non-associated flow response of Model-III can be specified
as
\begin{equation}
  \phi := \chi\big\vert_{\kappa=0}
        = \Norm{\BSigma}_{(a_1\nIP_1+a_2\nIP_2)}
        - \left[1 - \frac{\beta}{y_{12}}\right]
        + \frac{1}{2}{\Norm{\Bbeta}}^2_{(b_1\nIQ_1+b_2\nIQ_2)} \ .
 \label{flowpotential-2}
\end{equation}
Consequently, the evolution equations for a non-associated flow response
read
\begin{equation}
  \begin{array}{l@{\ }l@{\ }l}
    & \disp\dot{\STRAIN}^p &
    = \disp\lambda\left\{\frac{(a_1\nIP_1+ a_2\nIP_2):\BSigma}
           {\Norm{\BSigma}_{(a_1\nIP_1+a_2\nIP_2)}}\right\}\\[6.5mm]
    & \disp\dot{\Balpha} &
    = \disp\lambda\left\{\frac{(a_1\nIP_1+ a_2\nIP_2):\BSigma}
           {\Norm{\BSigma}_{(a_1\nIP_1+a_2\nIP_2)}}
    + (b_1\nIQ_1+ b_2\nIQ_2):\Bbeta\right\} \\[6.5mm]
    & \disp \dot{\alpha} &= \disp\lambda\frac{1}{y_{12}}
\end{array} \ .
  \label{evolution-M2-NA}
\end{equation}
\msectb{General remarks for algorithmic implementation.}
\label{AlgorithmicImplementation}
The next computational aspect is the time integration of the rate
equations of the models, subject to the constraint posed by the
respective yield conditions. The general return method suggested in \cite{simo+hughes2000,schroder+gruttmann+loblein2002}, together with a
backwards Cauchy-Euler integration scheme is used here in entire
analogy. While the parameters $\kappa,a_1,a_2$ and $y_{12}$ are obtained
from the experimental curves, the kinematic hardening parameters $b_1$
and $b_2$ are set to zero in the present work owing to absence of
relevant experimental data. A recipe for the identification of material
parameters associated with kinematic hardening can however be found in
\cite{voyiadjis+thiagarajan1996}. Furthermore, the evolution equations
of all the three models characterise Armstrong-Fredrick-type non-linear
kinematic hardening \cite{armstrong+frederick1996}, generalised to the
present case. For the choice $b = b_1 = b_2 = 0$, the models recover the
well known Melan-Prager-type kinematic hardening \cite{prager1956},
where $\Balpha$ is linear and homogeneous in $\dot\STRAIN^p$ \cite{papadopoulos+lu2001,miehe+apel+lambrecht2002}.

In summary, two different constitutive laws are implemented for each
model, namely
\begin{enumerate}
\item Model-I-a/Model-II-a/Model-III-a:~associative pressure-dependent
models, obtained by \reqs{general_energy} and \eqref{yield-1}/\eqref{plasticity_iso-MP2-_yield}/\eqref{yield-2}.
\item Model-I-b/Model-II-b/Model-III-b:~non-associative
pressure-dependent models, given by \reqs{general_energy},
\eqref{yield-1}/\eqref{plasticity_iso-MP2-_yield}/\eqref{yield-2} and
\eqref{flowpotential-1}/\eqref{plasticity_iso-MP2-_flow}/\eqref{flowpotential-2}.
\end{enumerate}

\msecta{Numerical simulations.}
\label{examples}
\begin{table}[t!]
  \begin{minipage}{\textwidth}
  \caption{Material parameters for Model-I.}
  \label{Table-1}
  \begin{tabular}{cllcll}
    No. & Name                                       & Par.            &        Value       & Unit \\\hline\hline
    1.  & Longitudinal Young's modulus               & $\EMODULUS_1$   & 130000             & $\mathrm{[MPa]}$ \\
    2.  & Transverse Young's modulus                 & $\EMODULUS_2$   & 11000              & $\mathrm{[MPa]}$ \\
    3.  & Longitudinal shear modulus                 & $\GMODULUS_{12}$ & 5800               & $\mathrm{[MPa]}$ \\
    4.  & Transverse shear modulus                   & $\GMODULUS_{23}$ & 3720               & $\mathrm{[MPa]}$ \\
    5.  & Poisson's ratio                            & $\POISSON_{12}$  & 0.306              & $\mathrm{[-]}$   \\
    6.  & Coefficient of hydrostatic pressure        & $\kappa$ & $0.9497\footnote{\label{note1}\textrm{Associative flow}}/1.105\footnote{\label{note2}\textrm{Non-associative flow}}$ & $\mathrm{[-]}$ \\
    7.  & Initial yield stress                       & $y_0$   & 10.6               & $\mathrm{[MPa]}$ \\
    8.  & Hardening modulus                          & $h$             & 237.9              & $\mathrm{[MPa]}$ \\
    9. & Pre-strain                                 & $\bar\alpha$    & 1$\times$10$^{-12}$ & $\mathrm{[-]}$   \\
    10. & Hardening exponent                         & $n$             & 0.249              & $\mathrm{[-]}$   \\\hline
  \end{tabular}
  \caption{Material parameters for Model-II.}
  \label{Table-2}
  \begin{tabular}{cllcll}
    No. & Name                                       & Par.            &        Value       & Unit \\\hline\hline
    1.  & Longitudinal Young's modulus           & $\EMODULUS_1$   & 130000                & $\mathrm{[MPa]}$ \\
    2.  & Transverse Young's modulus             & $\EMODULUS_2$   & 11000                 & $\mathrm{[MPa]}$ \\
    3.  & Longitudinal shear modulus             & $\GMODULUS_{12}$ & 5800                  & $\mathrm{[MPa]}$ \\
    4.  & Transverse shear modulus               & $\GMODULUS_{23}$ & 3720                  & $\mathrm{[MPa]}$ \\
    5.  & Poisson's ratio                        & $\POISSON_{12}$  & 0.306                 & $\mathrm{[-]}$   \\
    6.  & Anisotropic space yield strength       & $Y_{11}$         & $\approx\infty$       & $\mathrm{[MPa]}$ \\
    7.  & Anisotropic space yield strength       & $Y_{22}$         & 158.6                 & $\mathrm{[MPa]}$ \\
    8.  & Isotropic space yield strength         & $\bar{Y}$       & 158.6                 & $\mathrm{[MPa]}$ \\
    9.  & Coefficient of hydrostatic pressure    & $\kappa$        & $1.931\footref{note1}/1.917\footref{note2}$ & $\mathrm{[-]}$ \\
    10. & Initial yield stress                   & $y_0$           & 20.5                 & $\mathrm{[MPa]}$ \\
    11. & Hardening modulus                      & $h$             & 415.7                & $\mathrm{[MPa]}$ \\
    12. & Pre-strain                             & $\bar\alpha$    & 1$\times$10$^{-12}$   & $\mathrm{[-]}$   \\
    13. & Hardening exponent                     & $n$             & 0.241                & $\mathrm{[-]}$   \\\hline
  \end{tabular}
  \caption{Material parameters for Model-III}
  \label{Table-3}
  \begin{tabular}{cllcll}
    No. & Name                                   & Par.             & Value   & Unit \\\hline\hline
    1.  & Longitudinal Young's modulus           & $\EMODULUS_1$   & 130000             & $\mathrm{[MPa]}$ \\
    2.  & Transverse Young's modulus             & $\EMODULUS_2$   & 11000              & $\mathrm{[MPa]}$ \\
    3.  & Longitudinal shear modulus             & $\GMODULUS_{12}$ & 5800               & $\mathrm{[MPa]}$ \\
    4.  & Transverse shear modulus               & $\GMODULUS_{23}$ & 3720               & $\mathrm{[MPa]}$ \\
    5.  & Poisson's ratio                        & $\POISSON_{12}$  & 0.306              & $\mathrm{[-]}$   \\
    6.  & Transverse compressive yield stress    & $y_{22c}$        & $24.6\footref{note1}/27.4\footref{note2}$ & $\mathrm{[MPa]}$ \\[1mm]
    7.  & In-plane shear yield stress            & $y_{12}$         & 9.41               & $\mathrm{[MPa]}$ \\
    8.  & Transverse shear yield stress          & $y_{23}$         & 10.66              & $\mathrm{[MPa]}$ \\
    9.  & Hardening modulus                      & $h$             & 177.5              & $\mathrm{[MPa]}$ \\
    10. & Pre-strain                             & $\bar\alpha$    & 1$\times$10$^{-12}$ & $\mathrm{[-]}$   \\
    11. & Hardening exponent                     & $n$             & 0.246              & $\mathrm{[-]}$   \\\hline
  \end{tabular}
  \end{minipage}
\end{table}

The proposed models are implemented as user subroutines (UMAT) in
ABAQUS, a general purpose non-linear finite element program documented
in \cite{abaqus2013}. The subsequent numerical simulations demonstrate
the applicability and predictive capabilities of the models. In this
regard, the inelastic behaviour of a certain composite which has carbon
fibres reinforced in a polymer matrix (AS4/PEEK) is considered. The
pertinent experimental investigations are documented in \cite{vogler+kyriakides1999}. Note, that the simulations are conducted using a single
hexahedral 3D continuum element (C3D8). To avoid rigid body motions, the
bottom, left and rear faces are constrained in vertical, horizontal and
lateral directions, respectively, for the transverse compression load.
For the shear load, left face is fully constrained in all the directions
while the load is applied on the lateral face. The material parameters
used in the numerical simulations are listed in
\rTabs{Table-1}{Table-3}, wherein, the elastic parameters for all the
three models are taken directly from \cite{vogler+kyriakides1999,hsu+vogler+kyriakides1999}. Additionally, the elastic material parameters for
the fictitious isotropic space in Model-II are chosen to be
$\EMODULUS=4100$~MPa and $\POISSON=0.356$, which are the matrix
properties of PEEK \cite{vogler+kyriakides1999}. The plastic parameters
of Model-I and Model-II are calibrated using the procedure detailed in
our previous work \cite{nagaraja+pletz+schuecker2019}, from which all
the material parameters for Model-III are also taken.
\msectb{Calibration for the standard load cases.}
\label{calibration}
\inputfig[t!]{examples/calibration-results}{plot-1}
Figure \ref{plot-1} shows the calibration results comparing test data
\cite{vogler+kyriakides1999} and the three meso models for the two
standard test cases, namely the in-plane and transverse
compression test case. All the models reproduce the shear response and
the transverse compressive response accurately, as seen in
\rFig{plot-1}~(a) and (b). The experimental observations \cite{vogler+kyriakides1999} and computational verifications \cite{hsu+vogler+kyriakides1999,nagaraja+pletz+schuecker2019} are affirmative to the fact that the
constitutive response must be pressure sensitive to realistically
predict the non-linear behaviour of composite materials. Furthermore, it
should be noted from \rTabs{Table-1}{Table-3} that the coefficient of
hydrostatic pressure differs slightly between associative and
non-associative models.
\begin{table}
  \caption{Summary of the load paths}
    \label{Table-4}
    \begin{tabular}{lccccccl}
      Load path                                              & No.                                   & $\tau^\ast_{12}$ & -$\sigma^\ast_{22}$ & Unit \\\hline\hline
     Shear preload:~$\tau_{12}\rightarrow-\varepsilon_{22}$    & {\large \textcircled{\scriptsize 01}} & 43.1           & --                & [MPa] \\[1mm]
                                                             & {\large \textcircled{\scriptsize 02}} & 56.2           & --                & [MPa] \\[1mm]
                                                             & {\large \textcircled{\scriptsize 03}} & 66.9           & --                & [MPa] \\[1mm]
                                                             & {\large \textcircled{\scriptsize 04}} & 79.5           & 0                 & [MPa] \\[2mm]
     Compression preload:~$-\sigma_{22}\rightarrow\gamma_{12}$ & {\large \textcircled{\scriptsize 05}} & --             & 50.2              & [MPa] \\[1mm]
                                                             & {\large \textcircled{\scriptsize 06}} & --             & 84.83             & [MPa] \\[1mm]
                                                             & {\large \textcircled{\scriptsize 07}} & --             & 124.1             & [MPa] \\[1mm]
                                                             & {\large \textcircled{\scriptsize 08}} & --             & 164.5             & [MPa] \\[1mm]
                                                             & {\large \textcircled{\scriptsize 09}} & 0              & 242.6             & [MPa] \\\hline
    \end{tabular}
\end{table}

\msectb{Predictions for biaxial loads.}
\label{biaxial}
Next, the predictions of the three calibrated models are compared with
experimental results from the literature \cite{vogler+kyriakides1999}
for a set of biaxial loads. Two different load paths are considered
which are summarised in \rTab{Table-4}. For the first load path, referred to as the $\tau_{12}\rightarrow-\varepsilon_{22}$ path, the
specimen is first sheared to a predetermined stress level
($\tau^\ast_{12}$) and it is then compressed under displacement control
while the shear stress is kept constant. The final value of compressive
strain is chosen to be $-\varepsilon_{22}=4\%$. Likewise, in the second
load path, referred to as the $-\sigma_{22}\rightarrow\gamma_{12}$ path,
the specimen is first compressed to a desired stress level
($-\sigma^\ast_{22}$) and then sheared with $\gamma_{12}=4\%$ while keeping
the transverse compressive stress constant. The desired shear and
compressive stress values are taken directly from \cite{vogler+kyriakides1999}. Note that, the experimental investigations additionally examine
non-proportional load paths, where a set of load paths with proportional
increase of compressive and shear stresses is considered. A detailed
comparison of models predictions and experimental data for the
non-proportional load paths is documented in \cite{nagaraja+pletz+schuecker2019}, and hence not considered here.
\msectc{Predictions of the associated flow response.}
\label{shear/compression-preload-A}
\inputfig[t!]{examples/shear-preload-A}{plot-2}
\inputfig[b!]{examples/compression-preload-A}{plot-3}
Starting with the $\tau_{12}\rightarrow-\varepsilon_{22}$ load path,
\rFig{plot-2} shows a comparison of the model predictions with the
experimental results. The graphs in the left column show the compressive
responses under the shear preload along with a pure compression case.
The experimental results indicate that despite the initial yielding
caused by the shear preload, the compressive responses for load paths {\large\textcircled{\scriptsize 01}} and {\large\textcircled{\scriptsize 02}} are almost identical to {\large\textcircled{\scriptsize 09}}, while
that for {\large\textcircled{\scriptsize 03}} is somewhat lower. It can
also be inferred from the experimental results that the material
response first softens and then subsequently stiffens for increasing
shear preloads. The graphs in the right column show the increase in
shear strain caused by the compression load. It is seen that the
increase is comparatively large for load path {\large\textcircled{\scriptsize 03}} although a small bump in $\gamma_{12}$ can be seen at low
values of $\sigma_{22}$, which can be attributed to a test artefact
\cite{vogler+kyriakides1999}. The compressive responses predicted by
Model-I-a along with a pure compression case (load path {\large\textcircled{\scriptsize 09}}) are shown in \rFig{plot-2}~(a). The general
behaviour of softening and subsequent stiffening for increasing preloads
is not captured by the model. Additionally, for shear dominated stress
states (load paths {\large\textcircled{\scriptsize 01}}--{\large\textcircled{\scriptsize 03}}), erroneous predictions of Model-I-a can be seen
where tensile strain is predicted. Good qualitative agreement with the
experimental response is seen in \rFig{plot-2}~(b) with the predicted
shear strains being rather high. Figures \ref{plot-2}~(c) and (d) show a
comparison of Model-II-a predictions and experimental results for the
$\tau_{12}\rightarrow-\varepsilon_{22}$ load path. Though there are no
erroneous predictions for shear dominated loads, the general trends of
experimentally observed behaviour are not captured by the model, as seen
from \rFig{plot-2}~(c). For the load path {\large\textcircled{\scriptsize 03}}, there is an observable over-prediction by the model. In
comparison with the previous case, the agreement with the experimental
response is not satisfactory, as evident from \rFig{plot-2}~(d). Figures
\ref{plot-2}~(e) and (f) show the predictions of Model-III-a for the
$\tau_{12}\rightarrow-\varepsilon_{22}$ load path. The use of associative
flow rule induces physical inconsistencies in the response where tensile
transverse strains are predicted for shear dominated stress states (load
paths {\large\textcircled{\scriptsize 02}} and {\large\textcircled{\scriptsize 03}}). The predicted compressive response by the model agrees
well with experiments, as seen in \rFig{plot-2}~(e). Only for the
highest shear preload (load path {\large\textcircled{\scriptsize 03}}),
the response is slightly over-predicted. The general trend of the
experimental behaviour where the material response first shifts up and
then down for increasing shear preloads is also not captured by the
model, though the effect is minimal. The predicted shear strains agree
well with the experimental response (\rFig{plot-2}~(f)), except for load
path {\large\textcircled{\scriptsize 03}}. 

For the $-\sigma_{22}\rightarrow\gamma_{12}$ load path, results in a
similar format are shown in \rFig{plot-3}. The shear responses with
corresponding compression preloads are documented in the left column,
while the increase in compressive strain caused by shear load is shown
in the right column. One can generalise the insensitivity of the shear
response to the compressive preload as seen from the experimental data
in the left column. Only at the highest compressive preload ({\large\textcircled{\scriptsize 08}}), a small decrease in the shear stress is
seen. The insensitivity can also be inferred from plots in the right
column, where for load paths {\large\textcircled{\scriptsize 05}} and {\large\textcircled{\scriptsize 06}}) only a small increase in the
compressive strain is observed during shear load. Substantial increase
is only seen for load paths {\large\textcircled{\scriptsize 07}} and {\large\textcircled{\scriptsize 08}}). The predicted shear responses of
Model-I-a do not agree well with experiments as seen in
\rFig{plot-3}~(a). Owing to pressure-dependent plastic response
functions, the transverse compression load hinders the onset of yielding
and thus results in the reduced plastic flow \cite{hsu+vogler+kyriakides1999}. Consequently, the shear response in presence of compression is
over-predicted. Excellent agreement with experiments can be seen in
\rFig{plot-3}~(b). Likewise, the shear response in presence of
compression is over-predicted by Model-II-a in the
$-\sigma_{22}\rightarrow\gamma_{12}$ load path, as seen in
\rFig{plot-3}~(c). Good qualitative agreement is seen with the
experimental response in \rFig{plot-3}~(d). Predictions of Model-III-a
for the $-\sigma_{22}\rightarrow\gamma_{12}$ load path is shown in
\rFig{plot-3}~(e) and (f). Figure \ref{plot-3}~(e) depicts a comparison
of the model predictions and experimental shear response in presence of
compression, where a good conformity with experiments is seen. The
predicted compressive strains are in excellent agreement with
experiments which is evident from \rFig{plot-3}~(f).
\msectc{Erroneous predictions of the associated flow response.}
\label{PI-Asso}
\inputfig[t!]{examples/inconsistencies-corrections}{plot-4}
To illustrate erroneous predictions of the associated flow response,
the shear dominated $\tau_{12}\rightarrow-\varepsilon_{22}$ load path is
considered. The $\gamma_{12}$ vs.~$-\varepsilon_{22}$ strain response in
Figure \ref{plot-4}~(a), (c) and (e) shows Model-I-a, Model-II-a and
Model-III-a predictions for the load paths {\large\textcircled{\scriptsize 01}}--{\large\textcircled{\scriptsize 04}}. The assessment of the
plastic flow direction is apparent in these plots. For a just shear
stress state (load path {\large\textcircled{\scriptsize 04}}), positive
transverse strain is induced by Model-I-a and Model-III-a which is not
expected (\rFigs{plot-4}~(a) and (e)). The same phenomenon is observed
for shear dominated combined stress states (load paths {\large\textcircled{\scriptsize 01}}--{\large\textcircled{\scriptsize 03}}) where
physically unrealistic transverse tensile strain is predicted. This
aspect is also demonstrated in \cite{hsu+vogler+kyriakides1999} for load
paths {\large\textcircled{\scriptsize 05}}--{\large\textcircled{\scriptsize 08}} where the pressure-dependent model with an associative flow
rule predicts a decrease in the transverse strain for an increasing
shear strain, a trend opposite to the experimental results reported in
\cite{vogler+kyriakides1999}. For Model-II-a, it is seen that the
direction of plastic flow is generally aligned to the vertical axis for
just shear stress state, though a small amount of compressive strain is
induced (\rFig{plot-4}~(c)), again which is unrealistic. The
non-physical behaviour of associated flow response is also discussed in
the recent work \cite{laux+gan+barton+thomsen2019}, with respect to
compressive off-axis tests on a carbon-epoxy material. There, it is also
seen that tensile rather than compressive transverse strain is predicted
for $15^{\circ}$ and $45^{\circ}$ off-axis angles. This non-physical
behaviour is a consequence of the negative slope of the
Drucker-Prager-type yield surfaces used by these models, for just shear
and shear dominated combined stress states. The corrected material
response using a non-associative flow rule is shown in
\rFig{plot-4}~(b), (d) and (e). In line with the expectations, it is
seen that the direction of plastic flow is aligned to the vertical axis
for just shear stress state. Additionally, the predicted transverse
strains are compressive for shear dominated combined stress states. In
what follows, predictions of the non-associative flow rule are reported
and discussed based on the foregoing observations.
\msectc{Predictions of the non-associated flow response.}
\label{shear/compression-preload-NA}
\inputfig[t!]{examples/shear-preload-NA}{plot-5}
\inputfig[t!]{examples/compression-preload-NA}{plot-6}
Figure \ref{plot-5}~(a) and (b) show a comparison of Model-I-b
predictions and experimental responses for the
$\tau_{12}\rightarrow-\varepsilon_{22}$ load path. The non-associative
flow rule corrects the physically inconsistent material response
exhibited by the associative flow rule under shear dominated loads, as
seen in \rFig{plot-5}~(a). A good qualitative agreement with experiments
is seen for the compressive response. In comparison with Model-I-a, the
predicted shear strains are much lower, and significant deviations are
observed for load paths {\large\textcircled{\scriptsize 01}}--{\large\textcircled{\scriptsize 03}} which is visible from \rFig{plot-5}~(b).
Figure \ref{plot-5}~(c) and (d) shows the Model-II-b model predictions
for the $\tau_{12}\rightarrow-\varepsilon_{22}$ load path. Predictions of
the non-associative model are largely similar to that of the associative
model (Model-II-a). Figure \ref{plot-5}~(e) and (f) shows a comparison
of Model-III-b predictions and the experimental results for the
$\tau_{12}\rightarrow-\varepsilon_{22}$ load path. The predicted
compressive response by the model is in good agreement with experiments,
as seen in \rFig{plot-5}~(e). The predicted transverse strains are
compressive for the load paths {\large\textcircled{\scriptsize 01}}--{\large\textcircled{\scriptsize 04}}, thereby eliminating the
inconsistencies exhibited by Model-III-a. Additionally, there is no
over-prediction of transverse stress for load path {\large\textcircled{\scriptsize 03}}. Good agreement with the experimental response is seen
in \rFig{plot-5}~(f) in comparison with \rFigs{plot-5}~(b) and (d).

Figure \ref{plot-6} shows non-associative models predictions for the
$-\sigma_{22}\rightarrow\gamma_{12}$ load path. Similar to the associated
flow response, the shear response in presence of compression is
over-predicted by Model-I-b as seen in \rFig{plot-6}~(a). The
compressive strains are also over-predicted for load paths {\large\textcircled{\scriptsize 05}}--{\large\textcircled{\scriptsize 08}}. Figure
\ref{plot-6}~(c) and (d) shows Model-II-b predictions for the 
$-\sigma_{22}\rightarrow\gamma_{12}$ load path. Predictions of the
non-associative model are largely similar to that of the associative
model. Plots depicting Model-III-b predictions and experimental results
for the $-\sigma_{22}\rightarrow\gamma_{12}$ load path are shown in
\rFig{plot-6}~(e) and (f). The shear response in presence of compression
is slightly over-predicted by the model as seen in \rFig{plot-6}~(e).
This is a direct consequence of a higher value of the governing
coefficient of hydrostatic pressure, as higher transverse stress leads
to excessive stiffening in the shear response \cite{hsu+vogler+kyriakides1999}. The parameter $\kappa$ in \req{yield-2} should be rather low
based on the experimental results where shear responses of the load
paths {\large\textcircled{\scriptsize 04}}--{\large\textcircled{\scriptsize 07}} are almost the same, see \cite{vogler+kyriakides1999}. The
predicted $-\varepsilon_{22}$ agrees well with the experimental results
as seen in \rFig{plot-6}~(f) with slight overestimation for load path
{\large\textcircled{\scriptsize 08}}. 
\msectc{Discussion.}
\label{discussion}
An assessment of \rFigs{plot-2}, \ref{plot-3}, \ref{plot-5} and
\ref{plot-6} reveals that the predictions are in excellent agreement
with the experimental results whilst using an associative flow rule. For
the considered load paths, the compressive response with shear preload
and shear response with compression preload is over-predicted by all the
models, only at the highest value of the respective preload. It has been
reported in \cite{vogler+kyriakides1999} that the presence of shear
preload does not affect the compressive response significantly.
Likewise, the shear response is almost insensitive to the presence of
compression preload. These aspects are also reflected in the model
predictions. However, the associative flow rule exhibits non-physical
constitutive response under shear dominated combined stress states as
seen from \rFig{plot-4}. The non-associated flow response circumvents
these physical inconsistencies and yields the expected behaviour, but
notable deviations are observed for biaxial load paths where the shear
response in presence of compression is overestimated. A possible
explanation could be the influence of dilatation on the plastic
deformation due to crazing \cite{chen+gatea+ou+lu+long2016,zairi+abdelaziz+gloaguen+lefebvre2008}. In this case, the plastic flow potential
needs to be reformulated such that it is pressure dependent but stress
free in the fibre direction. A comprehensive discussion is beyond the
scope of the present work as it is unknown if crazing was observed in
the tests considered here. 

\msecta{Conclusions.}
\label{conclusions}
In this work, the effect of isotropic and anisotropic yield functions
in conjunction with associative and non-associative flow rules on the
non-linear inelastic behaviour of polymeric composites is investigated.
Three different plastic response functions (Model-I, Model-II and
Model-III) are considered. All the models are first calibrated to
reproduce the experimental pure shear and compression response. The
calibrated models are evaluated in detail by comparison to experimental
data for a range of bi-axial loads. The reported predictions show a high
degree of conformance with the experimental response.

The experimental investigations are affirmative to the fact that to
realistically predict the non-linear behaviour of polymeric composite
materials for different load combinations, the constitutive response
must be pressure sensitive. Further, plastic response functions that are
pressure-dependent but isotropic, need to be either mathematically
manipulated \cite{raghava+caddell+yeh1973,zhang+kikuchi+li+yee+nusholtz1998} or extended to anisotropic forms \cite{vogler+rolfes+camanho2013,nagaraja+pletz+schuecker2019} to reproduce the experimentally observed
biaxial response. It should also be emphasised here, that the models
based on the concept of mapped tensors \cite{car+oller+onate2000,car+oller+onate2001} do not fully ensure a linear elastic fibre response owing
to singularity problems of the transformation tensor. Clearly, the
stress tensor should be decomposed not just into volumetric and
deviatoric components, but also into the respective normal and shear
modes associated with the symmetry group. Only then can the experimental
biaxial response be captured accurately on the meso scale. Additionally,
it can be inferred that although both associative and non-associative
flow rules capture certain aspects of the polymeric composites, they do
not reproduce the full complexity. The non-associative flow rule
circumvents the physical inconsistencies induced by the associated flow
response under shear dominated combined stress states, and yields the
expected behaviour. However, notable deviations are observed for the
considered load paths where the shear response in presence of
compression is overestimated. This can be attributed to the choice of a
pressure-independent plastic flow potential (thereby a
pressure-independent flow rule) and a rate-independent setting, as a
result of which the predicted transverse strains are much higher than
those observed experimentally. At this point, the choice of a flow rule
is unclear and would require additional experimental data for bi-axial
responses, such as $\tau_{12}\rightarrow-\sigma_{22}$ which denotes the
evolution of yield surface and plastic flow potential. One can then
use the proposed associative or non-associative plasticity models for a
wide range of loading scenarios.

\def\path_lib{./}
\bibliographystyle{\path_lib/bib.bst}
\bibliography{\path_lib/Bibliography} 
\end{document}